\documentclass{cupconf}
\usepackage[dvips]{graphicx}

  \checkfont{eurm10}
  \iffontfound
    \IfFileExists{upmath.sty}
      {\typeout{^^JFound AMS Euler Roman fonts on the system,
                   using the 'upmath' package.^^J}%
       \usepackage{upmath}}
      {\typeout{^^JFound AMS Euler Roman fonts on the system, but you
                   dont seem to have the}%
       \typeout{'upmath' package installed. cupconf.cls can take advantage
                 of these fonts,^^Jif you use 'upmath' package.^^J}
      }
  \else
  \fi

  \checkfont{msam10}
  \iffontfound
    \IfFileExists{amssymb.sty}
      {\typeout{^^JFound AMS Symbol fonts on the system, using the
                'amssymb' package.^^J}%
       \usepackage{amssymb}%
         \let\leq=\leqslant
         \let\geq=\geqslant
      }{}
  \fi

  \IfFileExists{amsbsy.sty}
    {\typeout{^^JFound the 'amsbsy' package on the system, using it.^^J}%
     \usepackage{amsbsy}}
    {}

\newsavebox{\astrutbox}
\sbox{\astrutbox}{\rule[-5pt]{0pt}{20pt}}

\newcommand\eg{e.g.\ }

\newcommand{\be}{\begin{equation}}
\newcommand{\ba}{\begin{eqnarray}}
\newcommand{\ee}{\end{equation}}
\newcommand{\ea}{\end{eqnarray}}  
\newcommand{\etal}{et al.\ }
\def\gtsima{$\; \buildrel > \over \sim \;$}
\def\ltsima{$\; \buildrel < \over \sim \;$}
\def\gsim{\lower.5ex\hbox{\gtsima}}
\def\lsim{\lower.5ex\hbox{\ltsima}}
\def\simgt{\lower.5ex\hbox{\gtsima}}
\def\simlt{\lower.5ex\hbox{\ltsima}}
\def\simpr{\lower.5ex\hbox{\prosima}}

\def\sngg{PPSNe }
\def\snggo{PPSNe}
\def\msun{\,{\rm M_\odot}}

\def\E3{{\cal E}_{\rm g}^{III}}

\title[Pair-Production Supernovae]
{Pair-Production Supernovae: Theory and Observation}

\author[Evan Scannapieco]%
{Evan Scannapieco$^1$}
\affiliation{$^1$Kavli
Institute for Theoretical Physics, Kohn Hall, UC Santa Barbara, Santa
Barbara, CA 93106} 

\begin{document}

\maketitle

\begin{abstract}

Nonrotating stars that end their lives with masses $140 \msun \leq
M_\star \leq 260\,\msun$ should explode as pair-production supernovae
(\snggo).  Here I review the physical properties of these objects as
well as the prospects for them to be constrained observationally.  

In very massive stars, much of the pressure support comes from the
radiation field, meaning that they are loosely bound, and that $(d \lg
p/d \lg \rho)_{\rm adiabatic}$ near the center is  close to the
minimum value necessary for stability.  Near the end of $C/O$ burning,
the central temperature  increases to the point that photons begin to
be converted into electron-positron pairs, softening the equation of
state below this critical value.   The result is a runaway  collapse,
followed by  explosive burning that completely obliterates  the
loosely-bound star.  While these explosions can be up to 100 times
more energetic that core collapse and Type Ia supernovae, their peak
luminosities are only slightly greater.  However,
due both to copious Ni$^{56}$ production and hydrogen  recombination,
they are brighter much longer, and remain observable for $\approx$ 1
year.

Since metal enrichment is a local process, PPSNe should occur in
pockets of metal-free gas over a broad range of  redshifts, greatly
enhancing their detectability, and distributing their nucleosyntehtic
products about the Milky Way.   This means that measurements of the
abundances of metal-free stars should be thought of as directly
constraining these objects.  It also means that ongoing supernova
searches, which limit the contribution of very massive stars to $\lsim
1\,$\% of the total star formation rate density out to $z \approx 2,$
already provide weak constraints for PPSN models. A survey
with the NIRCam instrument on  JWST, on the other hand, would be able to 
extend these limits to $z \approx 10.$  
Observing a $0.3$ deg$^2$ patch of sky for $\approx 1$ week per year
for three consecutive years, such a program would
either detect or rule out the existence of these
remarkable objects.

\end{abstract}

\firstsection 

\section{Introduction}

Pair-production supernovae (PPSNe) are the uniquely calculable
result of nonrotating stars that end their lives in 
the $140-260 \msun$ mass range (Heger \& Woosley 2002, hereafter HW02).  
Their collapse and explosion result from an instability
that generally occurs whenever the central temperature
and density of star moves within a well-defined regime 
(Barkat, Rakavy, \& Sack 1967).
While this instability arises irrespective of the metallicity
of the progenitor star, PPSNe are in expected
only in primordial environment, and there are three main reasons
for this association.

Firstly, in the present metal-rich universe, it appears that stars
this large are never assembled, as supported by a wide range of
observations.  Figer (2005) carried out a detailed study of the $Z
\approx Z_\odot$ Arches cluster, which is large ($M_\star > 10^4
\msun$), young ($\tau$ = 2.0 - 2.5 Myrs), and at a well-determined
distance, making it ideal for such studies.  No stars more massive
than  130 $\msun$ were found, although more  than 18 were expected.
A similar $\approx 150 \msun$ limit was found in the lower metallicity
cluster R136 in the Large Magellanic Cloud (Weidener \& Kroupa 2003),
and from a  grab-bag of clusters compiled by Oey \& Clarke (2005).

However, there are good theoretical reasons to believe that the
situation may have been very different  under primordial
conditions. In this case the primary coolant at low temperatures is
molecular hydrogen, which starts to be populated according to
local thermodynamic equilibrium  (LTE) at a
typical density and temperature of $\approx 10^4$ cm$^{-3}$ and 100 K.
As the Jeans  mass under these conditions is  $\approx 10^3 \msun$
the fragmentation of primordial molecular clouds may have been 
biased towards the 
formation of stars with very high masses (Nakamura \&
Umemura 1999; Abel Bryan \& Norman 2000,  Schneider \etal 2002; Tan \&
McKee 2004).

Finally, as very massive, radiatively supported
stars are only loosely bound they tend to drive large winds.
However, these winds are primarily line-driven and scale
with metallicity as $Z^{1/2}$ or faster  (Kudritzki 2000; Vink \etal
2001; Kudritzki 2002).  As long as another mechanism did not
act to generate significant mass loss in primordial stars 
(\eg Smith \& Owocki 2006)  this raises the real possibility that 
they may have not only been born, but have ended the
lives in the mass range necessary to drive PPSNe.

This conference proceeding summarizes both the underlying physics and
the prospects for observation of these most powerful of astrophysical
explosions, and it is structured as follows: In \S2  I describe the
pair-production instability in detail, and how it eventually leads to
stellar disruption.  In \S3 I discuss the post-explosion physics of
these objects, and how it effects their luminosity and temperature
evolution. In \S4 I discuss the optical light-curves of PPSNe, 
and contrast them with  lightcuves of 
SNe Type-II and Ia.  \S5 discusses the redshift
evolution of metal-free stars, and its implications for PPSN
environments.  In \S6 these estimates are used to determine
the feasibility of present and future PPSN. In
\S 7 and \S 8 I discuss the likely distribution of the descendents of
metal-free stars in the Galaxy, and I close with a short summary in
\S9.

\section{Physics of Pair-Production Supernovae}

\begin{figure}
\centerline{\includegraphics[width=4.0in]{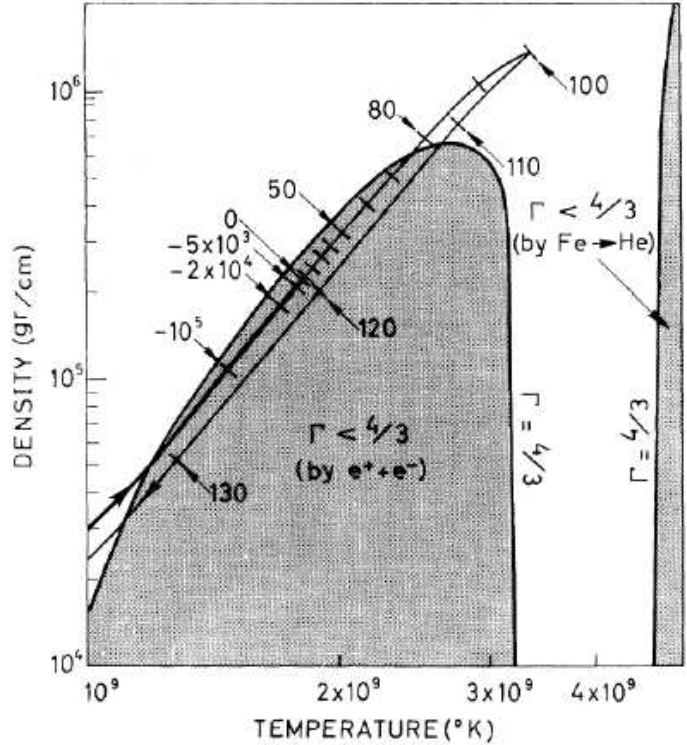}}
\caption{Original plot from Barkat, Rakavy, \& Sack (1967) showing the
range of central temperatures and densities at which the equation of
state softens below the $\gamma = 4/3$ value required for stability.
The solid lines give an early estimate of the evolution of a  star with
a 40 $\msun$ oxygen core near the end of its life.
Finally the region on the right shows the temperature and
density region relevant in usual core-collapse supernovae.}
\label{fig:instability}
\end{figure}

The instability that leads to the formation of PPSNe
was first identified by Barkat, Rakavy, \& Sack (1967),
who carried out a detailed analysis of the relevant equation of
state for very massive stars near the end of their lifetimes.  These
results are shown in Figure  \ref{fig:instability}.  In hydrostatic
balance $ p \propto M^{2/3} \rho^{4/3},$ which means that if the
adiabatic coefficient softens to below $\gamma = 4/3,$ the star will
become unstable to runaway collapse.

Figure \ref{fig:instability} shows that such a collapse occurs  for
stars with central temperatures from $10^9$ to $3 \times 10^9,$  and
central densities less than $5 \times10^5$ g cm$^{-1}.$  This region
is bounded by three limits.  Below the low-temperature boundary on the 
left, most of the central pressure is provided by the
radiation field, such that $\gamma \approx 4/3.$  As the temperature
increases, photon energies rise to the point that electron-positron
pairs begin to be made, removing energy from the radiation field and
softening $\gamma$ below the critical value.   The high-density end of this
region, on the other hand, is bounded by degeneracy pressure, which
provides sufficient support to halt collapse at  high densities.
Finally, at the high temperature boundary of the pair-instability regime, 
the energy  consumed in creating the rest mass becomes less significant,
and $\gamma$ remains above $4/3.$ 

As massive stars are very loosely
bound, their evolution is somewhat more complicated that their lower
mass counterparts, and thus requires more detailed numerical modeling.
As stars with masses above 140 $\msun$
collapse, their central temperature and density quickly
increases, moving right through the unstable regime and
starting explosive burning in the carbon-oxygen core.  This explosive
burning takes place in an environment in which there are very 
few excess neutrons, which results in a large deficiency in the 
number of nuclei with odd charge above $^{14}$N.   Thus
elements such as Na, Al, and P are almost 2 orders of magnitude 
less abundant than neighboring element with even charge such as
Ne, Mg, and Si.  This strong ``odd-even effect'' is a very general
feature that is unavoidable in a model of the nucleosynthetic
products of PPSNe (HW02).

Models of the kinematics of PPSNe are similarly robust.  For stars
with initial masses more than about 140 $\msun,$ the energy  released
during explosive burning is sufficient to completely disrupt the
star, resulting in a PPSN.  This complete disruption  means that
while the evolution of these stars is somewhat more complex than lower
mass stars, the explosion mechanism driving the resulting supernovae
is far simpler.  There are no issues of fallback, mass-cut, or
neutrino heating, and in the nonrotating case, the results are
uniquely calculable.

A shock moves outward from the edge of the core, initiating the
supernova outburst when it reaches the stellar surface.  Just above
the 140 $\msun$ limit, weak silicon burning occurs and only trace
amounts of radioactive $^{56}$Ni are produced and it is this $^{56}$Ni
that powers the late-time supernova light curves.  The amount of
$^{56}$Ni produced increases in larger progenitors, and in 260 $\msun$
progenitors up to 50
$\msun$ may be synthesized, $\approx$ 100 times more than in a typical
Type Ia supernova.  For stars with masses above 260 $\msun,$ however,
the onset of photodisintegration in the center imposes an additional
instability that collapses most of the star into a black hole (Bond,
Arnett \& Carr 1984; HW02; Fryer, Woosley, \& Heger 2001).

Pair-production supernovae are among most powerful thermonuclear
explosions in the universe, with a total energies ranging from $3
\times 10^{51}$ ergs for a 140 $\msun$ star (64 helium $\msun$ core)
to almost $100 \times 10^{51}$ ergs for a 260 $\msun$ star (133
$\msun$ helium core; HW02).  In Scannapieco \etal (2005, hereafter
S05), we used the implicit hydrodynamical code KEPLER (Weaver,
Zimmerman, \& Woosley 1978) to model the entire evolution of the star
and the resulting light curves.  KEPLER implements gray
diffusive radiation transport with approximate deposition of energy by
gamma rays from radioactive decay of $^{56}$Ni and $^{56}$Co (Eastman
\etal 1993), and the light curves obtained can only be followed as long as
there is a reasonably well-defined photosphere.

A supernova can be bright either because it makes a lot of radioactive
$^{56}$Ni (as in Type Ia supernovae) or because it has a large low
density envelope and large radius (as in bright Type II supernovae).
More radioactivity gives more energy at late times, while a larger
initial radius results in  a higher luminosity at early times.  Here,
the most  important factors in determining the resulting light curves
are the mass of the progenitor star and the efficiency of dredge-up of
carbon from the core into the hydrogen envelope during or at the end
of central helium burning.  The specifics of the physical process
encountered here are unique to primordial stars.  Lacking initial
metals, they have to produce the material for the CNO cycle
themselves, through the synthesis of $^{12}$C by the triple-alpha
process.  Just enough $^{12}$C is produced to initiate the CNO cycle
and bring it into equilibrium: a mass fraction of $10^{-9}$ when
central hydrogen burning starts, and a mass fraction $\sim 10^{-7}$
during hydrogen-shell burning.

At these low values, the entropy in the hydrogen shell remains barely
above that of the core, and the steep entropy gradient at the upper
edge of the helium core that is typical for metal-enriched
helium-burning stars is absent. This means that, during helium burning
the central convection zone  can get close, nip at, or even penetrate
the hydrogen-rich layers.  Once such mixing of high-temperature
hydrogen and carbon occurs, the two components burn violently, and
even without this rapid reaction, the hydrogen burning in the CNO
cycle increases proportionately to the additional carbon.  Thus mixing
of material from the helium-burning core, which has a carbon abundance
of order unity, is able to raise the energy generation rate in the
hydrogen-burning shell by orders of magnitude over its intrinsic value.

This mixing has two major effects on the PPSN progenitor: first, it
increases the opacity and energy generation in the envelope, leading
to a red-giant structure for the presupernova star, in which the
radius increases by over an order of magnitude.  Second, it decreases
the mass of the He core, consequently leading to a smaller mass of
$^{56}$Ni being synthesized and a smaller explosion energy.  The
former effect increases the luminosity of the supernova, at early
times, while the latter effect can weaken it, and in  S05 we accounted
for these uncertainties by employing different values of convective
overshooting.

\begin{table*}
\begin{center}
\caption{\footnotesize Properties of PPSN progenitor models}
\begin{tabular}{l|cccccc} \hline
Model &   $M_{\rm He}$ ($\msun$)  &  $M_{\rm N}$ ($\msun$) &
  $M_{^{56}{\rm Ni}}$ ($\msun$) &  $R$ ($10^{13}{\rm cm}$) &  ${\cal
  E}_{\rm kin}$ ($10^{51}{\rm ergs}$) \\ \hline\hline

150-W   &  70   &  3.5(-4)  &  4.2(-2)  & 3.9    &  6.9   \\ 
150-I   &  46   &  1.1(-4)  &  6.3(-2)  & 16     &  9.2   \\ 
150-S   &  49   &  0.86     &  8.6(-2)  & 26     &  8.5  \\ 
200-W    &  97   &  2.7(-6)  &   3.3    &  0.68  &  29.5  \\ 
200-I    &  58   &  8.0(-6)  &   5.1    &  2.8   &  36.5  \\ 
200-S    &  89   &  0.34     &   2.2    &  29    &  29.1  \\ 
200-S2   &  78   &  4.75     &   0.82   &  20    &  18.7  \\ 
250-W    &  123  &  3.1(-6)  &   6.2    &  0.58  &  47.2  \\ 
250-I    &  126  &  9.1(-6)  &   32     &  4.0   &  76.7  \\ 
250-S    &  113  &  1.34     &   24.5   &  26    &  64.6  \\ 
\hline
\end{tabular}
\end{center}
\end{table*}

A suite of representative models was chosen to address the expected
range of presupernova models - from blue supergiant progenitors with
little or no mixing to well-mixed red hypergiants, which can have
pre-SN radii of 20 AU or more.   These models are summarized in Table
1, in which the names refer to the mass of the progenitor star (in
units of $\msun$) and the  weak (W), intermediate (I), or strong (S)
level of of convective  overshoot.  Here we show the final mass in
helium, nitrogen, nickel, the radius just before the explosion, and
the kinetic energy of the explosion in  units of $10^{51}$ ergs.

\section{Luminosity and Temperature Evolution}

\begin{figure}
\centerline{\includegraphics[width=5in]{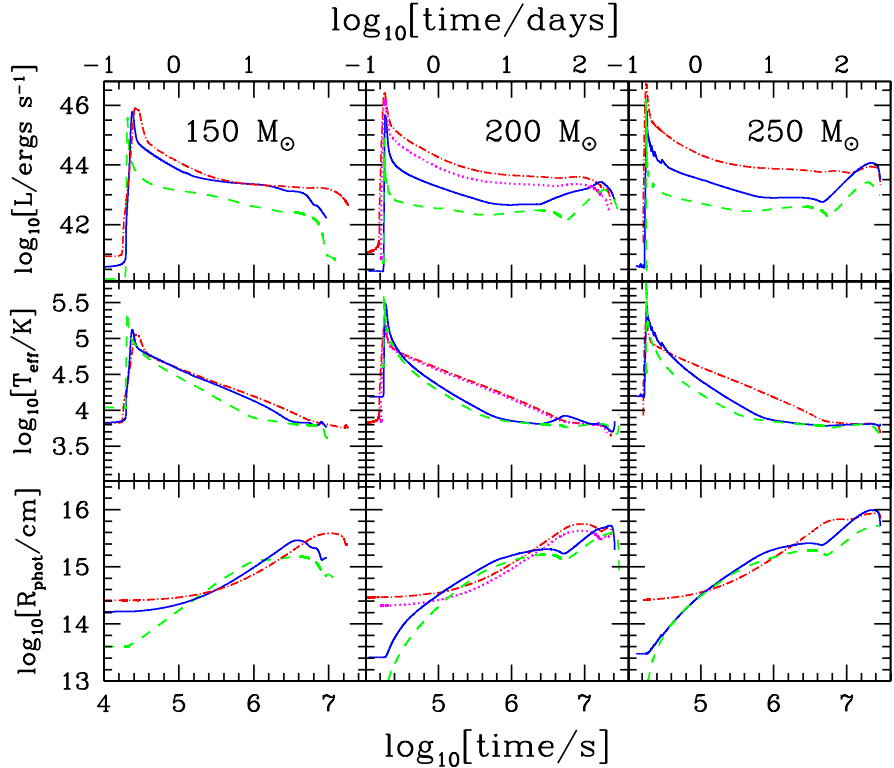}}
\caption{Luminosities (top row), effective temperatures (center row),
and photospheric radii (bottom row) of \sngg for ten different
representative models.   Each sets of panels are labeled by the total
mass of the progenitor star, and models with weak (dashed),
intermediate (solid) and  strong (dot-dashed) convective overshoot are
shown.  Finally, the dotted lines in the central panels correspond to
model 200-S2.  See text and Table 1 for details.}
\label{fig:sne}
\end{figure}

The KEPLER code can be used to compute approximate light curves and
has been validated both against much more complex and realistic codes
such as EDDINGTON and observations of a prototypical Type II-P
supernova, SN 1969L (Weaver and Woosley 1980; Eastman et al 1994),
although  its main deficiency is that it is a single temperature code
using flux-limited radiative diffusion. The evolution of the
luminosities, effective temperatures,  and photospheric radii for the
models in Table 1, are shown in Figure \ref{fig:sne}.  As the shock
moves toward the low-density stellar surface, its energy is deposited
into progressively smaller amounts of matter.  This results in high
velocities and temperatures when the shock reaches the stellar
surface, causing a pulse of ultraviolet radiation with a
characteristic timescale of a few minutes.  This ``breakout'' phase is
by far the most luminous and bluest phase of the PPSN burst, but its
very short duration makes it difficult to use in observational
searches.  In fact, the analog of this phase in conventional SNe has
so far only been indirectly detected in SN 1987A (\eg Hamuy \etal
1988; Catchpole \etal 1988).

Following breakout, the star expands with $R_{\rm phot}$ initially
proportional to time.  Though a small fraction of the outer mass may
move much faster, the characteristic velocity of the photosphere
during this phase is a modest $v = (2 KE/M)^{1/2} \sim (10^{53} {\rm
ergs}/200 \msun)^{1/2} \sim 5000$ km/s, because of the very large mass
participating in the explosion.  During the expansion, the
radiation-dominated ejecta  cool adiabatically, with $T$ approximately
proportional to $R^{-1},$ with an additional energy input from the
decay of $^{56}$Ni (if a significant mass was synthesized during the
explosion) and hydrogen recombinations (when $T \approx 10^4$K).  As
the scale radius for this cooling is the radius of the progenitor, the
temperatures and luminosities are substantially larger throughout this
phase in the cases with the strongest mixing.

After $\approx 50$ days, the energy input from $^{56}$Co decay becomes
larger than the remaining thermal energy and  the energy deposited by
$^{56}$Co in deeper layers that were enriched in $^{56}$Ni can diffuse
out. For stars that were compact to begin with, this can cause a
delayed rise to the peak of the light curve.  For stars with larger
radii, the radioactivity just makes a bright tail following the long
plateau in emission from the expanding envelope.  Eventually, even the
slow-moving inner layers recombine and there is no longer a
well-defined photosphere.  At this time the assumption of LTE breaks
down, and more detailed radiative transfer calculations are required,
which are beyond the scope of our S05 modeling. The SN is fainter and
redder during this phase, however, and thus is difficult to detect at
cosmological distances in optical and near infrared (NIR) surveys.

\section{PPSNe Lightcurves}				   

\begin{figure}
\centerline{\includegraphics[width=5in]{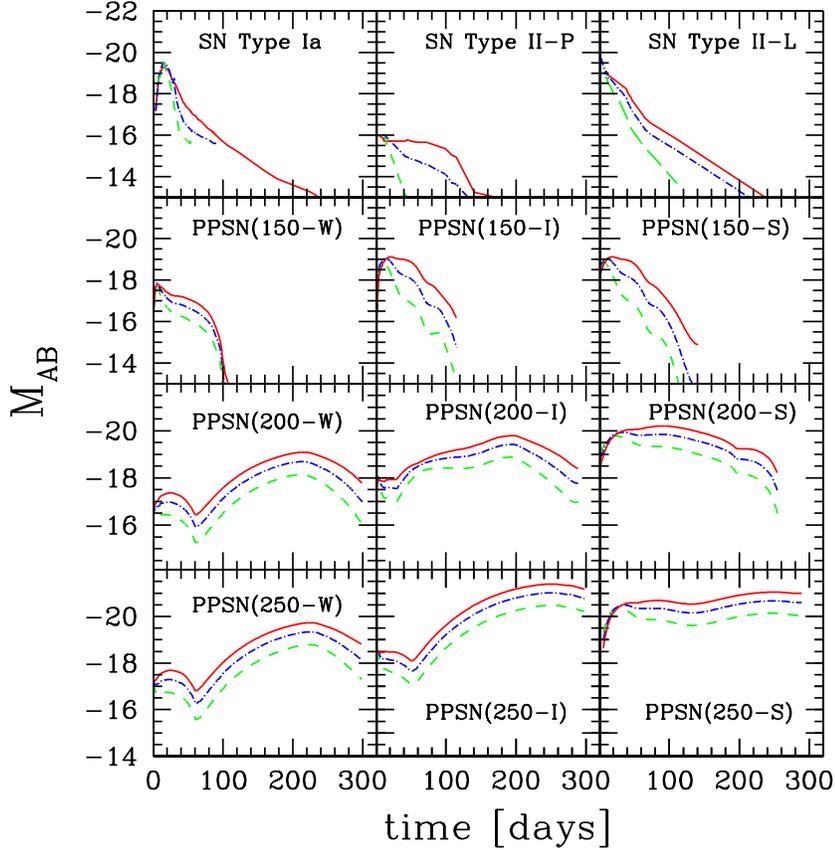}}
\caption{Comparison of light curves of a SN Type Ia, a SN Type II-P, a
bright SN Type II-L, and \sngg models with varying progenitor masses
and levels of dredge-up.  In all cases  the solid lines are absolute
V-band AB magnitudes, the dot-dashed lines are the absolute B-band AB
magnitudes, and  the dashed lines are the absolute U-band AB
magnitudes.}
\label{fig:lightcurves}
\end{figure}

In S05, we calculated approximate \sngg lightcurves, assuming a black body
distribution with the color temperature equal to the effective
temperature.  Recall that the peak frequency in this case occurs at  a
wavelength of $5100  (T_{\rm eff}/10000K)^{-1}$ \AA.    The
resulting light curves are shown in  Figure \ref{fig:lightcurves}
which gives the evolution of  AB magnitudes at three
representative wavelengths: 5500 \AA, corresponding to the central
wavelength of the V-band; 4400 \AA, corresponding to the B-band; and
3650 \AA, corresponding to the U-band.  We focus on blue wavelengths
as it is features in these bands that will be redshifted into the
optical and NIR at cosmological distances.  For comparison, we also
include observed light curves for a SN Type Ia  (1994D as measured 
by Patat \etal 1996 and Cappellaro \etal 1997) a SN Type II-P
(1999em as observed by Elmhamdi \etal 2003), and the very bright
Type II-L SN 1979C (de Vaucouleurs \etal 1981; Barbon \etal
1982).

The most striking feature from this comparison is that despite
enormous kinetic energies of  $\sim 50 \times 10^{51}$ ergs, the peak
optical luminosities of \sngg are similar to those of other SNe, even
falling below the Ia and II curves in many cases.  This is because the
higher ejecta mass produces a large optical depth and most of the
internal energy of the gas is converted into kinetic energy by
adiabatic expansion.  Furthermore, the colors of the  PPSN curves are
not unlike those of more usual cases.  In fact, pair-production
supernovae spend most of their lives in the same temperature range as
other SNe.  Clearly, then, \sngg will not be obviously distinguishable
from their more usual counterparts ``at first glance.''

Rather, distinguishing \sngg will from other SNe will require multiple
observations that constrain the time evolution of these objects. In
particular there are two key features that are uniquely characteristic
to \snggo.   The first of these is a dramatically extended intrinsic
decay time, which is especially noticeable in the models with the
strongest enrichment of CNO in the envelope. This is due to the long
adiabatic cooling times of supergiant progenitors, whose radii are
$\sim 20$ AU, but whose expansion velocities are similar or even less
than those of other SNe.  Second, \sngg are the only objects that show
an extremely late rise at times $\geq 100$ days.  This is due to
energy released by the decay of $^{56}$Co, which unlike in the Type Ia
case, takes months to dominate over the internal energy imparted by
the initial shock.  In this case the feature is strongest in models
with the least mixing and envelope enrichment during helium burning,
as these have the largest helium cores and consequently the largest
$^{56}$Ni masses.

Note, however, that neither of these features is generically present
in all \snggo, and both can be absent in smaller VMS that fail expand
to large sizes through dredge-up and do not synthesize appreciable
amounts of $^{56}$Ni.  In the 150-W case, for example, the luminosity
decays monotonically on a relatively short time scale, producing a
lightcurve similar to the comparison Type II curves.  In fact this
150 $\msun$ SN shares many features with its smaller-mass cousin: both
are SNe from progenitors with radii $\sim 10^{13}$ cm and in both
$^{56}$Ni plays a negligible role.  In no case, however, do \sngg look
anything like SNe Type Ia.  In particular none of the  pair-production
models display the long exponential decay seen in the Type Ia curves,
and all \sngg contain hydrogen lines, arising from their substantial
envelopes.

\section{The Redshift Evolution of Metal-Free Stars}

Planning searches for PPSNe not only depends on understanding their
lightcurves, but also the environments and redshifts at which they are
most like to be located. In Scannapieco, Schneider, \& Ferrara, (2003;
hereafter SSF03) we showed that  cosmological enrichment is a local
process, such  that the transition from metal-free to Population II
stars is heavily dependent on the efficiency with which metals where
mixed into the intergalactic medium.   This efficiency depends in turn
on the kinetic energy input from \snggo, which
was parameterized by the ``energy input per unit primordial gas mass''
${\cal E}_{\rm g}^{III}$, defined as  the product of the fraction of
gas in each primordial object that is converted into stars
($f_\star^{III}$), the number of \sngg per unit mass of metal-free
stars formed (${\cal N}^{\pm}$), the average kinetic energy per
supernova (${\cal E}_{\rm kin}$), and the fraction of
the total kinetic energy channeled into the resulting galaxy outflow
($f_{\rm wind})$.

Here we focus on the later stage of metal-free star formation.  Note
that this is fundamentally different than star frormation taking place
in very small ``minihalos'' at redshifts $\approx 25$,  which depends
sensitively on the presence of initial $H_2$  (\eg O'Shea \etal 2005).
In small objects, molecular hydrogen is easily photodissociated by
11.2-13.6 eV photons, to which the universe is otherwise transparent.
This means that the emission from the first stars quickly destroyed
all avenues for cooling by molecular line emission (Dekel \& Rees
1987; Haiman, Rees, \& Loeb 1997; Ciardi, Ferrara, \& Abel 2000),
which quickly raised the minimum virial temperature necessary to cool
effectively to approximately $10^4$ K. Thus the majority of primordial
star formation is likely to have occurred in objects above this limit,
who form their own $H_2$ at high densities and are largely impervious
to the photodissociating background (Oh \& Haiman 2002).

Incorporating outflows into a detailed analytical
model of such ``primordial galaxies'' leads to the approximate
relation that, by mass, the fraction of the total star formation in
metal-free stars at $z=4$ is  \be F_\star^{III}(z=4) \sim 10^{-5}
({\cal E}_{\rm g}^{III})^{-1},  \ee  where, as above,  ${\cal
E}^{\pm}$  is in units of $10^{51}$ ergs  per $\msun$ of gas (see
Figure 3 of SSF03 for details).   Extrapolating the results in SSF03
to $z=0$ gives  \be F_\star^{III}(z=0) \sim 10^{-5.5}  ({\cal
E}^{III}_{\rm g})^{-1}.  \ee  These fractions can be related to the
underlying population of stars by adopting fiducial values of
$f_\star^{III} = 0.1$ for the star formation efficiency, which is
consistent with the observed  star formation rate density at
intermediate and high redshifts (Scannapieco, Ferrara, \& Madau 2002),
and $f_w = 0.3$ for the wind efficiency,  which is consistent with the
dwarf galaxy outflow simulations  of Mori, Ferrara, \& Madau
(2002). Finally, we assume that  1 pair-production SN  occurs per 1000
solar masses of metal  free stars.  This gives $F_\star^{III}$ values
of  $0.3 ({\cal E}_{\rm kin})^{-1}$ at $z=4$ and  $0.1 ({\cal E}_{\rm
kin})^{-1}$ at $z=0,$ respectively.   Or, in other words, for typical
energies of $30 \times 10^{51}$ ergs per \snggo, $\sim 1\%$ of the
star formation at $z=4$ and  $\sim 0.3 \%$ of the star formation at
$z=0$ by mass should be in metal-free stars.

\begin{figure}
\centerline{\includegraphics[width=5in]{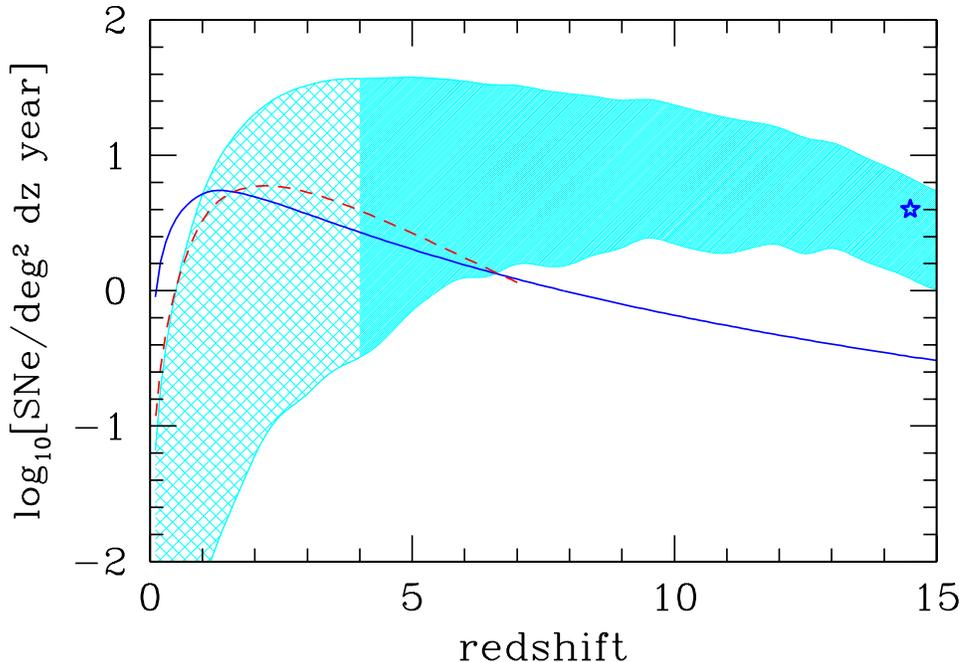}}
\caption{Number of \sngg per square degree per unit redshift  per year
for a wide range of models.  The solid and
dashed curves assume Pop III star formation rate densities of 0.001
$\msun$ yr$^{-1}$ Mpc$^{-3}$ and $1 \%$ of the observed star formation
rate density, respectively.    The shaded region covers  the range of
metal-free star formation  rate density models  considered in SSF03,
with the weakest feedback model (${\cal E}^{III}_{\rm g} = 10^{-4}$)
defining the upper end, and the strongest feedback model (${\cal
E}^{III}_{\rm g}= 10^{-2.5}$) defining the lower end.  An
extrapolation of these star formation rate densities to $z=0$ leads to
the crosshatched region.  In all SSF03 models the highest rates occur
at redshifts $\leq 10.$  Finally, the starred point is the $z=15$
estimate by Weinmann \& Lilly (2005).}
\label{fig:peryear}
\end{figure}

In Figure \ref{fig:peryear} we show estimates of the number of SNe per
deg$^2$ per dz {\em per year} over the wide range of models considered
in SSF03, extrapolating to $z=0.$  In all cases we assume that 1 PPSN
forms per 1000 solar masses of metal-free stars, and for comparison we
show two simple estimates which we refer to further below.   In the
first simple model, we  assume that metal-free  star  formation occurs
at a constant rate density, which we take to be  $\rho^{\pm} (z) 0.001
\, \msun \, {\rm yr}^{-1} \, {\rm Mpc}^{-3},$  independent of
redshift.   In the second case, we assume that  at all redshifts
metal-free stars form at 1\% of the observed total star formation rate
density, which we model as $\log_{10}[\rho^{\rm obs}_\star(z)/ \msun
\, {\rm yr}^{-1} \, {\rm Mpc}^{-3}]  = -2.1 + 3.9 \, \log_{10}(1+z)  -
3.0 \, [\log_{10}(1+z)]^2,$ a simple fit to the most recent
measurements (Giavalisco \etal 2004; Bouwens \etal 2004). For both
star formation models we again assume that  1 pair-production SN
occurs per 1000 solar masses of metal  free stars.

Note that for the full range of models in Figure \ref{fig:peryear},
metal-free star formation naturally occurs in the smallest galaxies,
which are just large enough to overcome the thermal pressure of the 
ionized IGM,
but small enough not to be clustered near areas of previous star
formation (SSF03).  In our adopted cosmology, for a temperature
of $10^4$K, the minimum virial mass is $3 \times 10^9 (1+z)^{-3/2}
\msun$ with a corresponding gas mass of $5 \times 10^8 (1+z)^{-3/2}
\msun.$ This means the total stellar mass of primordial objects is
likely to be around $M_\star \approx 10^8 \msun (1+z)^{-3/2}$, many
orders of magnitude below $L_\star$ galaxies.  Thus in general
blank-field surveys should be the best method for searching for
\snggo, as catalogs of likely host galaxies would be extremely
difficult to construct.

Nevertheless, as VMS shine so brightly, a direct search for primordial
host galaxies is not a hopeless endeavor.  In particularly the lack of
dust in these objects and the large number of ionizing photons from
massive metal-free  stars leads naturally to a greatly enhanced Lyman
alpha luminosity.  Following SSF03 this can be estimated as   \be
L_\alpha = c_L (1-f_{\rm esc}) Q(H) M_\star, \ee where $c_L \equiv
1.04 \times 10^{-11}$ ergs, $f_{\rm esc}$ is the escape  fraction of
ionizing photons from the galaxy, which is likely to be $\lesssim 0.2$
(see Ciardi, Bianchi, \& Ferrara 2002 and references  therein), and
the ionizing photon rate $Q(H)$ can be estimated as $\approx 10^{48}$
s$^{-1}$ $\msun^{-1}$ (Schaerer 2002).  This gives a value of
$L_\alpha  \sim 10^{45} (1+z)^{-3/2} {\rm ergs \, s}^{-1}$  which, if
observed in a typical $~1000$ \AA \, wide broad band corresponds to an
absolute AB mag $\sim -23 + 3.8 \log(1+z),$ much brighter than the
\sngg themselves.  However this flux would be spread out over many
pixels and be more difficult to observe against the sky than the
point-like \sngg emission.  For further details on the detectability of 
metal-free stars  though Lyman-alpha observations, the reader is 
referred to SSF03.

\section{Pair-Production Supernovae in Cosmological Surveys}

From the models developed above, it is relatively straightforward to
relate the star formation history of VMS to the resulting number of
observable pair-production supernova.  In this section and below we
adopt cosmological parameters of $h=0.7$, $\Omega_m$ = 0.3,
$\Omega_\Lambda$ = 0.7, and $\Omega_b = 0.045$, where $h$ is the
Hubble constant in units of 100 km s$^{-1}$ and $\Omega_m$,
$\Omega_\Lambda$, and $\Omega_b$  are the total matter, vacuum, and
bayonic densities  in units of the critical density (\eg Spergel \etal
2003).

Here we focus on three PPSN light curves, which bracket the range of
possibilities: the faintest of all our models, 150-W,  in which there
is neither significant dredge-up nor $^{56}$Ni production; an
intermediate model, 200-I, in which some dredge-up occurred, but 5.1
$\msun$ of $^{56}$Ni were formed; and the model with the brightest
lightcurves, 250-S, in which substantial dredge-up leads to an
enormous initial radius of over 20 A.U.,  and the production of 24.5
$\msun$ of $^{56}$Ni causes an extended late-time period of high
luminosity.  Note that we do not address the possibility of extinction
by dust, which amounts to assuming that pristine regions remain
dust-free thought the lifetime of the very massive PPSN progenitors
stars.

For any given PPSN model, we can calculate $t(\lambda,
F_\nu^{\rm min}, z)$  the total time the observed flux at the
wavelength $\lambda$  from a SN at the redshift $z$ is greater than
the magnitude  limit associated with the  specific flux $F_\nu^{\rm
min}.$  The total number of PPSNe  shining at
any given time with fluxes above $F_\nu^{\rm min}$, per square degree
per unit redshift is then given by the product of the volume element, the
(time-dilated) PPSN rate density, and the time a given PPSN is
visible, that is
\be
\frac{dN_{\rm deg^2}}{dz}(\lambda,F_\nu^{\rm min},z)  = [r(z) \,
\sin(1 {\rm deg})]^2 \, \frac{dr}{dz} \frac{\rho^{\pm}(z)}{1+z}
t(\lambda, F_\nu^{\rm min}, z),    
\ee  
where the rate density $\rho^{\pm}(z)$ is the
number of \sngg  per unit time per comoving volume as a function of
redshift.

\begin{figure}
\centerline{\includegraphics[width=5in]{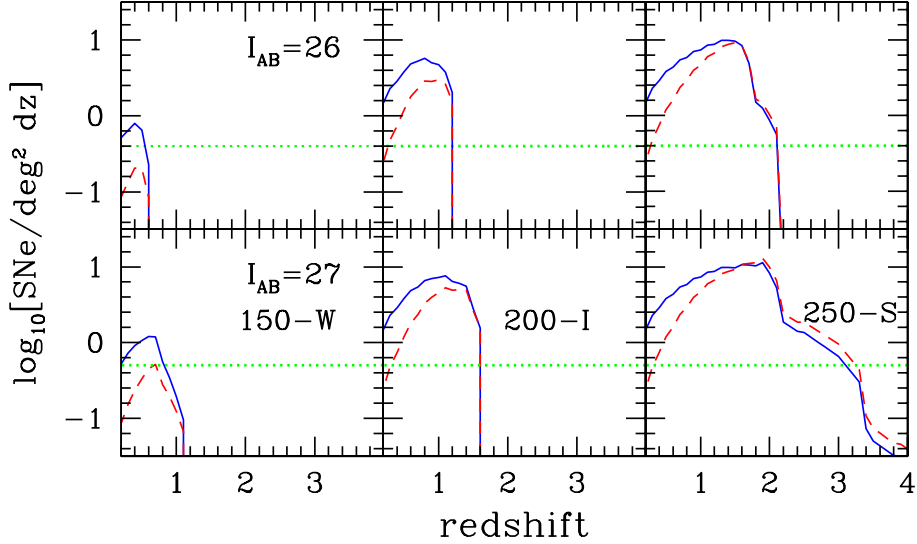}}
\caption{Number of \sngg  per square degree per unit redshift above a
given I-band magnitude, assuming 0.001 $\msun$ yr$^{-1}$ Mpc$^{-3}$
(solid lines) or 1\% of the observed star formation rate density (dashed
lines).   The $I_{AB} = 26$ cut taken in  the upper rows approximately
corresponds to the magnitude limit of the  Institute for Astronomy
Deep Survey which covered 2.5 deg$^2$ (as shown by the dotted lines)
from September 2001 to April 2002.    The $I_{AB} = 27$ limit in the
bottom panel corresponds to that of the ongoing COSMOS survey, which 
will survey an area of 2 deg$^2$ (again indicated by the dotted lines).
Note however, that the COSMOS survey itself is
primarily focused on large-scale structure issues and will not be
able to find PPSNe, as each pointing is visited only once.}
\label{fig:Iband}
\end{figure}

The resulting observed \sngg counts for these models 
are given in Figure \ref{fig:Iband} for two
limiting magnitudes.  In the upper panels, we take a $I_{\rm AB} = 26$
magnitude limit, appropriate for the Institute for Astronomy (IfA)
Deep Survey (Barris \etal 2004), a  ground-based survey that covered a
total of 2.5 deg$^2$ from September  2001 to April 2002.  
As we are focused on lower-redshift observations, we only
plot estimated counts for the two simple rate density models
described in \S5, rather than the more detailed (but higher redshift)
SSF03 models.


From this figure we see that existing data sets, if properly analyzed,
are easily able to place useful constraints on PPSN formation at low
redshifts.  Given a typical model like 200-I for example, the
already realized IfA survey can be used to  place a constraint   of
$\lesssim 1 \%$ of the total star formation  rate density out to a
redshift $\approx 1$.  Similarly, extreme models such as 250-S can be
probed out to redshifts $\approx 2$, all within the context of a
recent SN search driven by completely  different science goals.  Note
however that these limits are strongly dependent on significant mixing
in the SN progenitor or the production of $^{56}$Ni, and thus models
such  as 150-W remain largely unconstrained by the IfA survey.

\begin{figure}
\centerline{\includegraphics[width=5in]{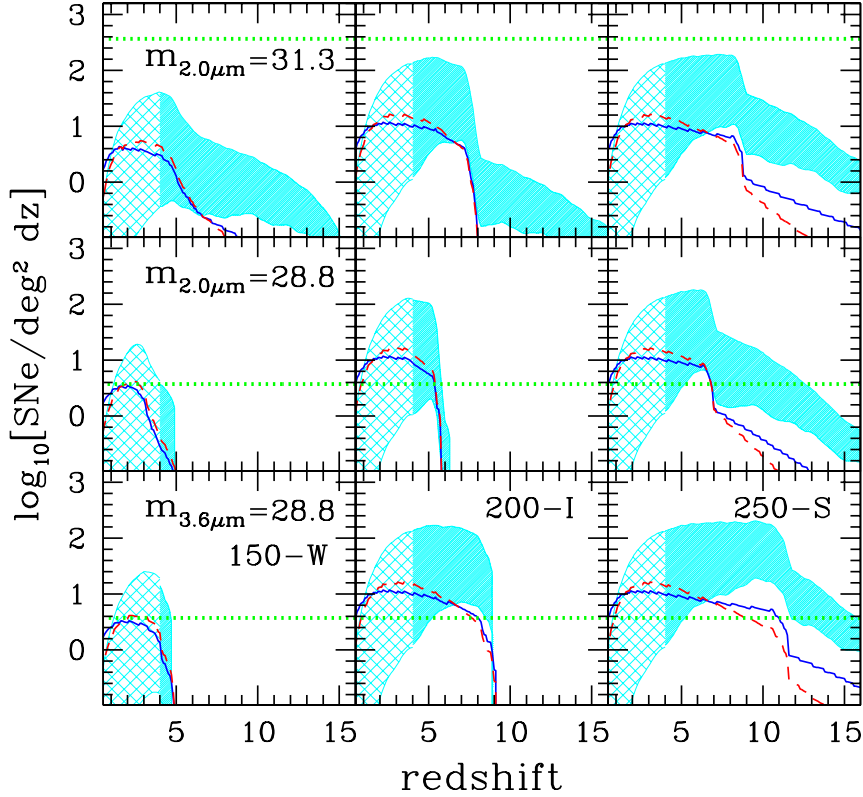}}
\caption{Number of \sngg  per square degree per unit redshift above a
fixed AB magnitude, with limits appropriate for three possible surveys
with the James Webb Space Telescope.  Curves are as in Figure
\protect\ref{fig:Iband}, while the shaded regions are  predictions
from the full range of SSF03 models described in \S 5.     {\em Top:}
The $31.1$ AB magnitude limit taken in these panels corresponds to the
10 $\sigma$ detection limit for a $10^6$ sec integration with at  $2
\mu m$ with the NIRCam instrument.  While this would be able to probe
SN extremely deeply, it would only cover a $9.8$ arcmin$^2$ patch of
the sky, corresponding to the dotted lines.  No detections would be
expected in this small area. {\em Center:} The $28.8$ AB magnitude $2
\mu m$ limit  taken in these panels  corresponds to the 10 $\sigma$
detection limit for a $10^4$ second integration with NIRCam.   In
$10^6$ seconds, such measurements could be taken at roughly 100
pointings, covering a $0.3$ deg$^2$ area, corresponding to the dotted
lines.  In this case $\sim$ 10 PPSNe like 150W or 200I  would be
detectable out to $z \approx 6$, while  up to $\sim 30$ PPSNe like
250S would be detectable out to $z\approx 10.$ {\em Bottom:} The
$28.8$ AB magnitude $3.6 \mu$ limit taken in these panels again
corresponds to 10 $\sigma$ detections for a $10^4$ second NIRCam
integration.   Moving to a longer wavelength results in a significant
boost to both the number of detectable PPSNe  and the maximum redshift
at which they can  be observed.}
\label{fig:Hband}
\end{figure}

In the bottom panels of Figure \ref{fig:Iband} we consider a limiting
magnitude of  $I_{\rm AB} = 27$, appropriate for the COSMOS
survey\footnote{see http://www.astro.caltech.edu/$\sim$cosmos/},  a
project that covers 2 deg$^2$ using the {\em Advanced Camera for
Surveys} on HST.  Raising the limiting magnitude from $I_{\rm AB} =
26$ to $I_{\rm AB} = 27$ has the primary effect of extending the
sensitivity out to slightly higher redshifts.  This pushes the probed
range from $z \lesssim 1$ to $z \lesssim 1.5$ in the 200-I case and
from  $z \lesssim 2$ to $z \lesssim 3$ in the 250-S case.  Again this
is all in the context of an ongoing survey.    Even with this fainter
limiting magnitude, however, low-luminosity \sngg like 150-W  are
extremely difficult to find, and remain largely unconstrained.

This shortcoming is easily overcome by moving to NIR wavelengths.
In Figure \ref{fig:Hband} we present for the first time similar limits
computed for the James Webb Space Telecope  (JWST),using both the
simple models PopIII SFR density models shown in Figure
\ref{fig:Iband} as well as the full range of more detailed models
acomputed in SSF03.  With JWST, ${dN_{\rm deg^2}}/{dz}$ is
dramatically increased  with respect to ground-based searches.   This
is due to the fact that for the majority of their lifetimes, the
effective temperatures of \sngg are just above the $\approx 10^{3.8}
K$ recombination temperature of hydrogen, which corresponds to a peak
black-body wavelength  $\approx 8000$ \AA.   This means that for all
but the lowest redshifts, the majority of the emitted light is shifted
substantially redward of the I-band, which is centered at $9000$ \AA.
Given the sensitivities described in Gardner \etal (2006), a 10$^6$
second NIRCam at $2.0 \mu$  integration would be able to see a PPSN
out to $z \approx 15.$

In fact, NIRCam will be so sensitive that using it to perform
extremely long integrations will {\em not} be the best  way to search
for PPSNe.  Rather, the key issue will be covering enough area to find
them at the rates expected from theoretical models.  As a single
NIRCam pointing only covers $9.8$ square arcminutes, any individual
such field is not likely to host a PPSN, even at such very faint
magnitudes.  Rather a much more efficient method is to carry out a
survey composed of roughly 100 pointings, each with a $10^4$ second
integration time.   In this case,  $\sim$ 10 PPSNe like 150W or 200I
would be detectable out to $z \approx 6$, while  up to $\sim 30$ PPSNe
like 250S would be detectable out to $z\approx 10.$   Furthermore,
moving to slightly longer wavelengths, increases the high-redshift
sensitivity  much in the same way as  carrying out longer integrations
at a fixed band.   Thus the expected number of PPSNe/deg$^2/dz$ above
an AB magnitude limit of 28.8 at $3.6 \mu$m is comparable to the
number of PPSNe/deg$^2/dz$ above an  AB magnitude limit of 31.3 at
$2.0 \mu$m.

Taken together, these results imply that the optimal strategy for
searching for PPSNe with JWST will be to carry out a 2 or 3 band
NIRCam survey (to obtain color information on these objects), with an
emphasis on longer wavelengths (to boost sensitivity), made up of
$\approx 100$ pointings with moderate integration times (to maximize
sky coverage).   Finally as most of the features in PPSNe lightcurves
are on the 30-100 day scale, this field should be revisited
roughly once per $30$ days  $\times (1+z) \approx$   1 year, on
three occasions.    Although this would require about 1-2
weeks of dedicated time each year, clearly this program could be
carried out in the context of a more general deep-field study, such as 
the present Supernova Cosmology Project in the context of the Great 
Observatories Origins Deep Survey with HST.

\section{Modeling the Galactic  Descendants of Metal-Free Stars}

A secondary method of searching for PPSN is the detection of their
nucleosynthetic products in present day stars.   However,
interpretation of these abundances is complicated by the fact that
searches for metal-poor stars are limited to the Galactic halo, where
dust extinction and crowding are minimal.  In fact, these analyses have
already  provided a number intriguing constraints on  the enrichment
history of the halo  (Freeman \& Bland-Hawthorn 2002; Beers \&
Christlieb 2005), including uncovering the presence of extremely
heavy-element deficient stars (Christlieb \etal 2002; Frebel \etal
2005).  Yet, it is still unclear how the observed population of halo
stars is related to PopIII star formation  (White \& Springel 2000;
Diemand, Madau, \& Moore 2005).

To quantify this, in Scannapieco \etal (2006), we  combined a
high-resolution N-body simulation of the formation of the Milky-Way
with a semi-analytical model of metal enrichment similar to that
in SSF03.  Our N-body simulation was carried out with the
{\tt GCD+} code (Kawata \& Gibson 2003a), and it used a
multi-resolution technique (Kawata \& Gibson 2003b) to achieve
high-resolution within a 1 Mpc radius, while the outer regions
exerting tidal forces were handled with lower resolution.   In the
high-resolution region, the dark metter particle mass was $M_{\rm
vir}=7.8\times10^5$ ${\rm M}_\odot$, compared to the final virial mass
of  $M_{\rm vir}=7.7\times10^{11}$ ${\rm M}_\odot.$ The simulation
data was output every $0.11$ Gyr, and at each output, we use a
friend-of-friends (FOF) group finder to identify  the virialized DM
halos, with a linking parameter of $b=0.2.$ As in \S5 we then assumed
star formation to occur in all halos with virial temperatures above
the atomic cooling limit of $10^4$ K, which corresponds to a minimum
mass of $M_{\rm min} \equiv 3.0 \times 10^9 (1+z)^{-3/2} M_\odot.$
This means that even at a very high redshift of 20,  all the halos
relevant to our study contained at least 50 particles, and are
well-identified by a FOF group finder.

Our next step was to use this accretion history to identify two types
of objects: i) halos that collapsed out of primordial gas, which we
identified as PopIII objects containing ``the first stars,'' and
ii)  halos that collapsed from gas that has been enriched purely by
PopIII objects, which we identified as second-generation objects
objects containing  ``the second stars.''  Following
our approach in SSF03, we adopted a model of outflows as spherical
shells expanding into the Hubble flow (Ostriker \& McKee 1988)  for
both PopIII and PopII/I objects.  These shells were assumed to be
driven only by the internal hot gas pressure and decelerated by
accreting material and gravitational drag.   
The evolution of each such bubble was then
completely determined by the mechanical energy imparted to the
outflow.  In particular the only difference between
PopIII and PopII/I outflows arises from the energy input per unit gas mass,''
${\cal E}_{\rm g}^{III,II}$.

In the PopII/I case, we calculated this assuming that $10\%$ of the
gas was converted into stars, that $10^{51}$ ergs of kinetic energy
input 300 $\msun$ of stars formed, and we assumed an overall wind
efficiency fit to results of Mori, Ferrara, \& Madau (2002) and
Ferrara, Pettini, \& Shchekinov (2000).  In the
PopIII case, on the other hand, as there are no direct constraints, we
varied $\E3$ over a large range as in \S 5.  Finally, when outflows
slowed down to the IGM sound speed, we assumed they fragmented and let them 
expand with the Hubble flow.
For further details, the reader is referred to Scannapieco \etal
(2006).

\begin{figure}
\centerline{\includegraphics[width=5.3in]{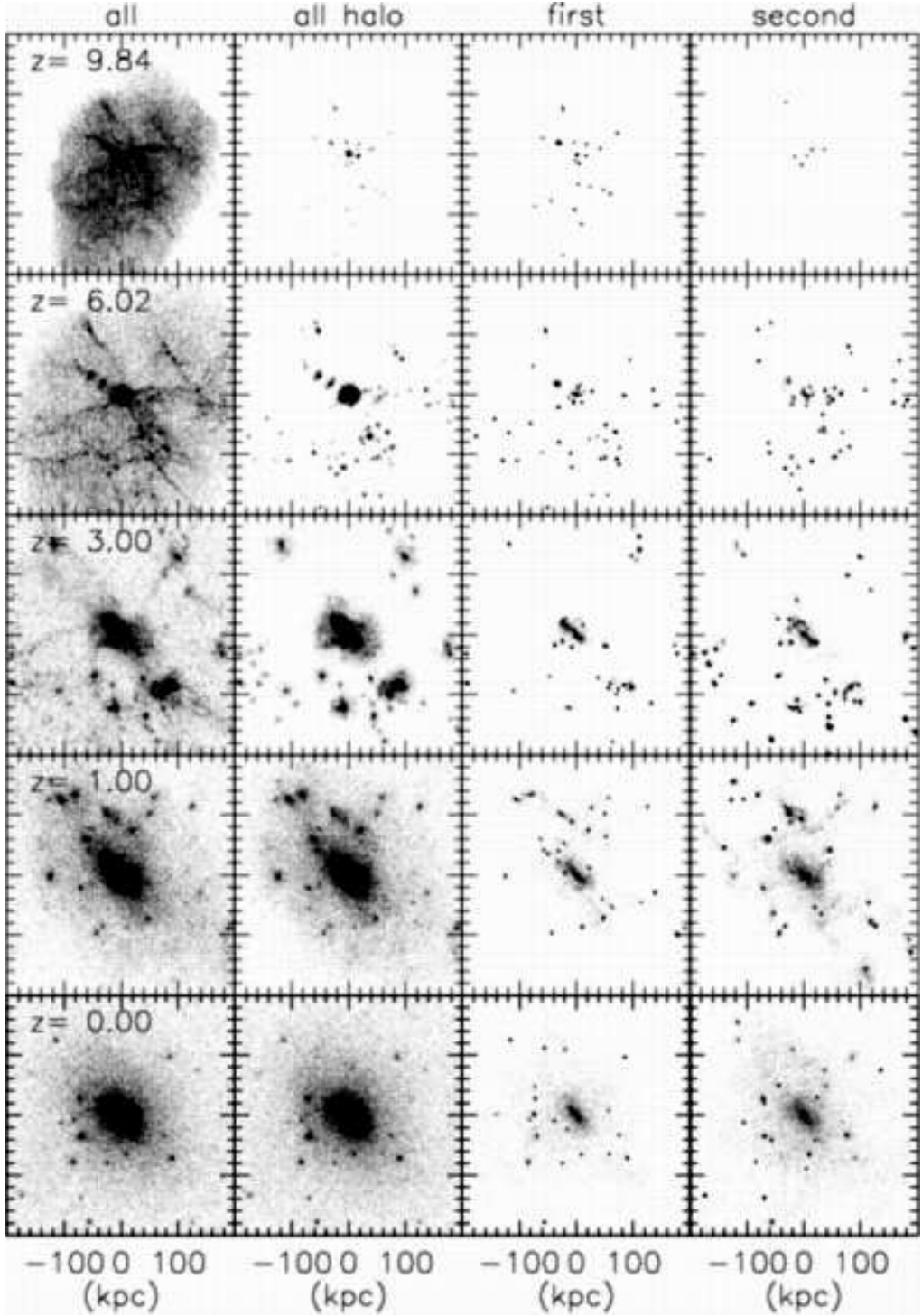}}
\caption{Distributions of first and second stars  (3rd and 4th columns) 
in a a 200 proper kpc$^3$ region, at various redshifts, 
for a model with $\E3 = 10^{-2}$.
For comparison, the 1st and 2nd columns show all the particles and
all the particles that were ever within a halo with a virial
temperature above the atomic cooling limit, respectively.}
\label{fig:anim}
\end{figure}

\section{The Spatial Distribution of the Galactic 
  Descendants of Metal-Free Stars}

Figure \ref{fig:anim} demonstrates the positions of the first and
second stars at different
redshifts in the Lagrangian model with $\E3 = 10^{-2}$.  
At early times ($z=9.84$), the first stars form close to the
central density peak of the progenitor galaxy, due
to the higher density peaks in this region.   Second stars form in
the halo in the neighborhood of first stars,  because they condense
from gas that is enriched by the material from the explosions of the
first stars.  At a later time ($z=6.02$), new first stars are still
forming, but now on the outer regions of the progenitor galaxy, because they
are not yet affected by winds from the central region.  
The formation of the first and second
stars is complete around $z=3$, at which time the full region is
enriched with metals. Also at this redshift, first and second stars
start to be accreted into the assembling Galactic halo.   This
assembly is almost complete by $z=1$, and thus the distribution at
this redshift is similar to that at $z=0$.

\begin{figure}
\centerline{\includegraphics[width=5in]{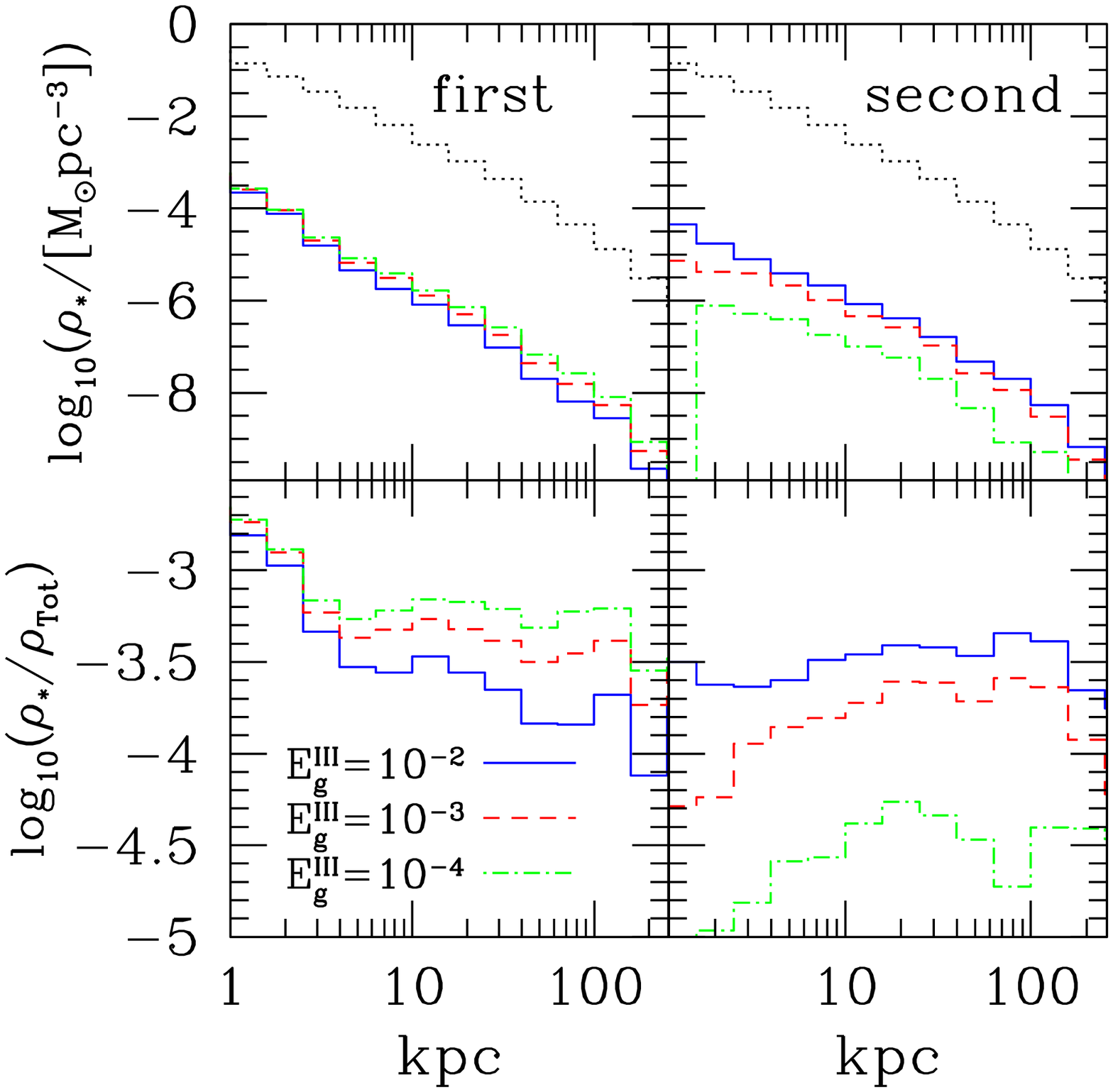}}
\caption{Radial profile of  first (left column) and second stars
(right column) at $z=0$.  In the top
row, the dotted lines give the overall dark matter profile of the
galaxy, which is compared with the radial density of first and second
stars in  models with $\E3 = 10^{-2}$ (solid), $\E3 = 10^{-3}$
(short-dashed), and and $\E3 = 10^{-4}$ (dot-dashed).
The lower row shows the fraction of the total density
in first and second stars, with symbols as in the upper panels.
Note that this plot does not include any mass loss due to stellar 
evolution.}
\label{fig:radial1}
\end{figure}

In Figure \ref{fig:radial1} we plot the radial mass density of first
and second stars, as compared to the dark matter distribution in our
simulation.  In computing these masses, we assume that 10\% of the gas
mass  in each halo is converted into stars  and we intentionally make
no attempt to account for mass loss due to stellar evolution.
Furthermore, we plot only out to 250 kpc, which is 1/4 the size  of
the high-resolution region.  Here we see that the density profiles of
first and second stars are similar to the total dark matter density
profile, although the second stars have a slightly shallower slope. As
a result, the  density of first stars at the center is 100 times
higher than at the 8 kpc orbital radius of the sun.  However, the
important number for developing observational strategies for finding
the first stars is their relative density with respect to the field
stars.  While this is not directly computed in our simulation, the
lower panels of  Figure \ref{fig:radial1} show the  local mass density
of stars normalized by the local  density of dark matter.

Amazingly, the mass fraction contained  in the first stars varies only
very weakly with radius.  Moving from 1 to 100 kpc in the  $\E3 =
10^{-4}$ model, for example,  the fraction of the mass in first stars
decreases only by a factor $\approx 4.$ Furthermore, increasing the
efficiency of PopIII winds to $\E3 = 10^{-2}$ has the effect of
decreasing the fraction of first stars without strongly affecting
their radial distribution.  

This is both good and bad news for PPSNe models.  First of all it
means that  observations of metal-poor stars in the Galactic halo
should be taken as directly constraining the properties of primordial
objects.    Thus the lack of metal-free stars observed in the halo is
likely to imply a real lower mass limit in the metal free initial mass
function (IMF) of at least $0.8 \msun,$ the mass of a low-metallicity
star  with a lifetime comparable to the Hubble time (Fagotto \etal
1994).  In fact the $\sim 10^{-6} M_\odot$ pc$^{-3}$ value for PopIII
stars in the solar neighborhood we compute is so high as compared to
the observed stellar mass density of $5 \times 10^{-5} M_\odot$
pc$^{-3}$ (Preston, Shectman, \& Beers 1991), that several primordial
$0.8 \msun$ stars would have been observed even if $f_*^{III}$ were
over an order of magnitude lower than the $0.1$ value we used to
normalize our approach.  This lends strong support to models of metal
free stars as biased to high masses.

On the other hand, it also means that the abundances we see in very
metal  poor stars should be taken to constrain elements produced in
the first stars.   In particular, if a large fraction of PopIII star
formation resulted in PPSNe, then then the strong odd-even effect
discussed in \S2  should be measurable in in a subset of stars  with
[Fe/H] $\leq -2$.  To date these measurements  have failed to uncover
this signal.  As discussed in Tumlinson (2006), this does not rule out
the presence of PPSNe, but it does argue strongly that a significant
number of  PopIII stars ended their lives as more usual core-collapse
supernovae.

\section{Summary}

Theoretically, PPSNe are the simplest of all supernovae.  Driven by a
well-understood dynamical instability, and leading to complete stellar
disruption, they are the uniquely  calculable result of nonrotating
stars than end their lives in  the $140-260 \msun$ mass range (Heger
\& Woosley 2002).   The issue is only when and where such stars
existed.  In the present, enriched, universe the observed upper mass
limit  of  forming stars, and the rate of mass loss in O stars argue
strongly against these objects.  However, in the primordial
high-redshift universe things are likely to have been very different.
The typical fragmentation mass under these conditions is $\approx 1000
M_\odot$ and stellar winds, at least of the line driven type observed
today, are expected to be negligible.   This raises the real
possibility that in the metal-free universe a large fraction of stars
generated PPSNe.

As metal-enrichment is an intrinsically local process that proceeds
over an extended redshift range, at each redshift, pockets of
metal-free star formation are naturally confined to the lowest-mass
galaxies, which are small enough not to be clustered near areas  of
previous star formation.  As such faint galaxies are difficult to detect
and even more difficult to confirm as metal-free, the hosts of \sngg
could easily be lurking at the limits of present-day galaxy surveys.

In S05 we showed that the most important factorsfor modeling PPSNe lightcurves
are the
mass of the progenitor star and the efficiency of dredge-up of carbon
from the core into the envelope.  In general, increasing the mass
leads to greater $^{56}$Ni production,  which boots the late time SN
luminosity.   Mixing, on the other hand, has two major effects: it
leads to a red giant phase that increases the early-time SN luminosity
and it decreases the mass of the He core, consequently leading to a
somewhat smaller mass of $^{56}$Ni.  Despite these uncertainties,
\sngg in general can be characterized by: (1) peak
magnitudes that are brighter than Type II SNe and comparable or
slightly brighter than typical SNe Type Ia;  (2) very long decay times
$\sim 1$ year, which result from the  large initial radii and large
masses of material involved in the explosion;  and (3)  the presence
of hydrogen lines, which are caused by the  outer envelope.

The S05 lightcurves also allowed us to calculate
the number of \sngg detectable in current
and planned supernova searches.  Here the long lifetimes help to keep
a substantial number of \sngg  visible at any given time, meaning that
ongoing SN searches  should be able to limit the contribution of VMS
to  $\lsim 1\,$\% of  the total star formation rate density out to a
redshift of 2, unless both mixing and $^{56}$Ni production are absent
for all \snggo. Such constraints already place meaningful  limits on
the  cosmological models.

The impact of future NIR searches is even more promising, as the
majority of the PPSN light is emitted at restframe wavelengths
longward of $\approx 8000$ \AA.  In particular JWST surveys with
NIRCam have the potential to  place fantastic constraints on PPSNe out
to  $z \approx 10.$  In this case the best approach will be a $\approx
0.3$ deg$^2$ survey made up of $\approx 100$ NIRCam pointings with
$\approx 10^4$ sec integrations in two or three bands, with emphasis
on the redder colors.     Furthermore this field should be revisited
with a cadence of roughly once per year on  three occasions.
Although  this would require about 1-2 weeks of dedicated time each
year, clearly this program could be carried out in the context of a more
general deep field study, with a much broader set of science goals.

Closer to home, we have also studied the final distribution of the
elements synthesized in primordial stars.   Despite the large
uncertainties involved, all models generically predict
significant PopIII star formation in what is now the Galactic
halo. Thus, if they   have sufficiently long lifetimes, a significant
number of stars formed in  initially primordial star clusters should
be found in ongoing surveys for metal-poor halo stars. This is both good
and bad news for PPSNe models.  While it implies a real lower mass
limit for the PopIII IMF  of at least $0.8 \msun,$ it also suggests
that the lack of an odd-even effect in the observed  abundance ratios
of metal-poor stars should be taken as evidence that a significant
number of PopIII stars ended their lives as more usual core-collapse
supernovae.

However, definitive limits on PPSNe will only come from space based
NIR  surveys.  If they result in detections, they will open
a new window on star formation and the history of cosmological
chemical enrichment.  If they result in upper limits, they
will place exquisite constraints on the presence of PPSN forming in
primordial environments above the atomic cooing limit, but leave open
the question of their formation in the first ``minihalos'' collapsing
at extremely high redshift. While this would remain a possibility, the
absence of detections from space-based searches would limit PPSN to
the most remote and undetectable corner of the universe.  Undaunted
theorists might still wish to discuss them in workshops on the deepest
depths of the cosmological dark ages.  Observers may be reminded that
``he who wishes to lie, should put the evidence far away'' (Livio 2006).

\begin{acknowledgments}

I would like to thank my collaborators Chris Brook, Andrea Ferrara, 
Brad Gibson, Alexander Heger, Daisuke Kawata,  Piero Madau, Raffaella 
Schneider, 
and Stan Woosley, for allowing me to present the results of our work
together here.   I am also thankful to Jonathan Gardner, for providing
detailed information on the planned capabilities of the James Webb
Space Telescope.  Finally, I  would like to thank Mario Livio, Massimo
Stiavelli, and the other members of the organizing committee as well as
the many excellent speakers for a fun and informative symposium.

\end{acknowledgments}

\end{document}